\documentclass[intlimits,twoside,a4paper]{article}
\usepackage{amsmath,amssymb}
\usepackage{graphicx}
\usepackage{wrapfig}
\usepackage[T2A]{fontenc}
\usepackage[cp1251]{inputenc}

\usepackage{cmpj2}

\issue{2012}{15}{4}{43002}

\doinumber{10.5488/CMP.15.43002}

\title[Spin alternating chain in arbitrarily oriented field]
{Ferrimagnetic spin-1/2 chain of alternating
Ising \\ and Heisenberg spins in arbitrarily oriented \\ magnetic field}

\author[J.~Stre\v{c}ka  \textsl{et al.}]{J.~Stre\v{c}ka\refaddr{label1}, M.~Hagiwara\refaddr{label2},
Y.~Han\refaddr{label2}, T.~Kida\refaddr{label2}, Z.~Honda\refaddr{label3}, M.~Ikeda\refaddr{label2}}
\addresses{
\addr{label1} Department of Theoretical Physics and Astrophysics, Faculty of Science,
P.J. \v{S}af\'{a}rik University, \\ Park Angelinum 9, 040 01 Ko\v{s}ice, Slovak Republic
\addr{label2} KYOKUGEN (Center for Quantum Science and Technology under Extreme Conditions),
Osaka University, \\ 1-3 Machikaneyama, Toyonaka, Osaka 560--8531, Japan
\addr{label3} Department of Functional Materials Science, Graduate School of Science and Engineering, \\
Saitama University, 255 Simookubo Saitama, Saitama 338--8570, Japan}

\date{Received July 3, 2012}

\authorcopyright{J.~Stre\v{c}ka, M.~Hagiwara, Y.~Han, T.~Kida, Z.~Honda, M.~Ikeda, 2012}

\begin{document}

\maketitle

\begin{abstract}
The ferrimagnetic spin-1/2 chain composed of alternating Ising and Heisenberg spins in an arbitrarily oriented magnetic field is exactly solved using the spin-rotation transformation and the transfer-matrix method. It is shown that the low-temperature magnetization process depends basically on a spatial orientation of the magnetic field. A sharp stepwise magnetization curve with a marked intermediate plateau, which emerges for the magnetic field applied along the easy-axis direction of the Ising spins, becomes smoother and the intermediate plateau shrinks if the external field is tilted from the easy-axis direction. The magnetization curve of a polycrystalline system is also calculated by performing powder averaging of the derived magnetization formula. The proposed spin-chain model brings an insight into high-field magnetization data of $3d$-$4f$ bimetallic polymeric compound Dy(NO$_3$)(DMSO)$_2$Cu(opba)(DMSO)$_2$, which
provides an interesting experimental realization of the ferrimagnetic chain composed of two different
but regularly alternating spin-1/2 magnetic ions Dy$^{3+}$ and Cu$^{2+}$ that are reasonably approximated by the notion of Ising and Heisenberg spins, respectively.

\keywords ferrimagnetic spin chain, exact results, magnetization plateau, $3d$-$4f$ bimetallic compound
\pacs 05.30.-d, 05.50.+q, 75.10.Hk, 75.10.Jm, 75.10.Pq, 75.40.Cx
\end{abstract}

\section{Introduction}
\label{intro}

Exactly solved quantum spin chains belong to the most attracting issues to deal with in the condensed matter theory, because they are capable of providing a deeper understanding into many unconvential quantum cooperative phenomena~\cite{mat93}. Recently, the particular research interest has been turned towards sophisticated Ising-Heisenberg chains, which are aimed at describing hybrid spin systems composed of `classical' Ising and quantum Heisenberg spins~\cite{val08,oha09,ant09,can09,roj11,ana12,str05,heu10,sah12}. Among other matters, the Ising-Heisenberg chains have become helpful in providing the evidence for several novel and unexpected quantum states~\cite{val08,oha09,ant09,can09,roj11}, fractional magnetization plateaus in the low-temperature magnetization process~\cite{ant09,can09,roj11}, enhanced magnetocaloric effect during the adiabatic demagnetization~\cite{can09}, thermal entanglement~\cite{ana12}, etc. It is, therefore, quite challenging to search for suitable experimental realizations of the Ising-Heisenberg chains testifying to the aforementioned theoretical findings, but only a few experimental systems satisfy a very specific requirement
of a regular alternation of the Ising and Heisenberg spins. Up to now, the magnetic behaviour of three different polymeric chains Cu(3-Clpy)$_2$(N$_3$)$_2$~\cite{str05}, [(CuL)$_2$Dy][Mo(CN)$_8$]~\cite{heu10} and [Fe(H$_2$O)(L)][Nb(CN)$_8$][Fe(L)]~\cite{sah12} was successfully interpreted within the framework of the Ising-Heisen\-berg chains.

The main goal of the present work is to examine the magnetization process in the spin-$\frac{1}{2}$ chain of alternating Ising and Heisenberg spins, which brings an insight
into ferrimagnetism of $3d$-$4f$ bimetallic coordination compound Dy(NO$_3$)(DMSO)$_2$Cu(opba)(DMSO)$_2$~\cite{cal08}. The organization of this paper is as follows. Exact results for the total and sublattice magnetizations of the investigated spin-chain model are derived in section~\ref{sac}. The most interesting theoretical results are presented in section~\ref{result}, where they are also confronted with the relevant experimental data. The paper ends up with several concluding remarks given in section~\ref{conclusion}.

\section{Spin alternating chain}
\label{sac}

Let us consider the spin-$\frac{1}{2}$ chain composed of alternating Ising and Heisenberg spins in an external magnetic field of arbitrary spatial direction. It is quite plausible to suppose that magnetic properties of this 'classical-quantum' spin-chain model will be highly anisotropic due to the presence of the Ising spins. In this respect, it is of particular interest to examine how the magnetization process depends on a spatial orientation of the external magnetic field, which can be unambiguously given by the deviation angle $\theta$ referred with respect to a unique easy axis of the Ising spins. The investigated spin-chain model can be defined through the following Hamiltonian
\begin{eqnarray}
{\cal H} = - J \sum_{i=1}^{N} S_i^z (\sigma_{i}^z + \sigma_{i+1}^z) - g_1^z \mu_{\rm B} B \cos \theta \sum_{i=1}^{N} \sigma_i^z
- g_2^x \mu_{\rm B} B \sin \theta \sum_{i=1}^{N} S_i^x  - g_2^z \mu_{\rm B} B \cos \theta \sum_{i=1}^{N} S_i^z \, .
\label{ham}
\end{eqnarray}
Here, $\sigma_i^z$ and $S_i^{\alpha}$ ($\alpha = x,z$) denote standard spatial components of the spin-$\frac{1}{2}$ operator, whereas the former (latter) operators apparently refer to the Ising (Heisenberg) spins. The first summation takes into account the Ising-type exchange interaction $J$ between the nearest-neighbour Heisenberg and Ising spins, the second term determines the Zeeman's energy of the Ising spins in the external magnetic field with the projection $B \cos \theta$ towards their easy ($z$) axis, while the last two Zeeman's terms determine the overall magnetostatic energy of the Heisenberg spins affected both by the transverse ($B \sin \theta$) and longitudinal ($B \cos \theta$) component of the external magnetic field. The quantities $g_1^{\alpha}$ and $g_2^{\alpha}$ ($\alpha = x,z$) label spatial components of Land\'e $g$-factors of the Ising and Heisenberg spins, respectively, $\mu_{\rm B}$ is Bohr magneton and $B$ stands for the external magnetic field. Notice that the Hamiltonian~(\ref{ham}) is built on the assumption that the transverse component of Land\'e $g$-factor of the Ising spins is negligible ($g_1^{x} \approx 0$), i.e., the situation, which is realistic only for the highly anisotropic (the so-called Ising-type) magnetic ions~\cite{jon74,wol00}.

Taking advantage of a `classical' nature of the Ising spins, which represent a barrier for local quantum fluctuations induced by the transverse component of the external field acting on the Heisenberg spins, one may rewrite the total Hamiltonian~(\ref{ham}) as a sum of commuting site Hamiltonians
\begin{eqnarray}
{\cal H} = \sum_{i=1}^{N} {\cal H}_i \, ,
\label{hamsum}
\end{eqnarray}
whereas each site Hamiltonians ${\cal H}_i$ involves all the interaction terms of the $i$th Heisenberg spin and the Zeeman's energy of its two nearest-neighbour Ising spins
\begin{eqnarray}
{\cal H}_i = - J S_i^z (\sigma_{i}^z + \sigma_{i+1}^z) - H_2^x S_i^x - H_2^z S_i^z - \frac{H_1^z}{2} \left(\sigma_{i}^z + \sigma_{i+1}^z \right).
\label{hami}
\end{eqnarray}
For simplicity, we have introduced here three `effective fields' $H_1^z = g_1^z \mu_{\rm B} B \cos \theta$, $H_2^x = g_2^x \mu_{\rm B} B \sin \theta$, and $H_2^z = g_2^z \mu_{\rm B} B \cos \theta$ in order to write the Hamiltonian~(\ref{hami}) and all subsequent expressions in a more abbreviated form. Owing to the validity of the commutation relation $[{\cal H}_i, {\cal H}_j] = 0$ between different site Hamiltonians, the partition function of the considered spin-chain model can be partially factorized into the following product
\begin{eqnarray}
{\cal Z} = \sum_{\{ \sigma_i \}} \prod_{i=1}^{N} \mbox{Tr}_{S_i} \exp \left(- \beta {\cal H}_i \right)
         = \sum_{\{ \sigma_i \}} \prod_{i=1}^{N} \mbox{T}\,(\sigma_{i}^z, \sigma_{i+1}^z),
\label{pf}
\end{eqnarray}
where $\beta = 1/(k_{\rm B} T)$, $k_{\rm B}$ is Boltzmann's constant, $T$ is the absolute temperature, the symbol $\mbox{Tr}_{S_i}$ denotes a trace over
two spin states of the $i$th Heisenberg spin and the summation $\sum_{\{ \sigma_i \}}$ runs over all available configurations of the Ising spins.
To proceed further with a calculation, the partial trace over spin degrees of freedom of the Heisenberg spins must be performed before summing
over spin states of the Ising spins. It is, therefore, quite convenient to diagonalize the site Hamiltonian~(\ref{hami}) by making use of the
spin-rotation transformation
\begin{eqnarray}
S_i^x  = S_i^{x}{'} \cos \phi_i + S_i^z{'} \sin \phi_i \, , \qquad S_i^z  = -S_i^{x}{'} \sin \phi_i + S_i^{z}{'} \cos \phi_i \, ,
\label{rt}
\end{eqnarray}
which brings the site Hamiltonian~(\ref{hami}) into the diagonal form
\begin{eqnarray}
{\cal H}_i{\!'} = - S_i^{z}{'} \sqrt{\left[J \left(\sigma_{i}^z + \sigma_{i+1}^z\right)  + H_2^z \right]^2 + \left(H_2^x\right)^2}  - \frac{H_1^z}{2} \left(\sigma_{i}^z + \sigma_{i+1}^z \right)
\label{hamid}
\end{eqnarray}
provided that the spin rotation~(\ref{rt}) is performed by the specific angle
\begin{eqnarray}
\phi_i = \arctan \left [\frac{H_2^x}{J \left(\sigma_{i}^z + \sigma_{i+1}^z\right) + H_2^z} \right ].
\label{ra}
\end{eqnarray}

The effective Boltzmann's weight, which enters the factorized form of the partition function~(\ref{pf}), can be now simply evaluated
by employing the trace invariance and the diagonalized form of the site Hamiltonian (\ref{hamid})
\begin{eqnarray}
\mbox{T}\,(\sigma_{i}^z, \sigma_{i+1}^z) \!\!\!&=&\!\!\! \mbox{Tr}_{S_i} \exp \left(- \beta {\cal H}_i \right)
= \mbox{Tr}_{S_i{\!\!'}} \exp \left(- \beta {\cal H}_i{\!'} \right) \nonumber \\
\!\!\!&=&\!\!\! 2 \exp \left[\frac{\beta}{2} H_1^z \left( \sigma_{i}^z + \sigma_{i+1}^z \right) \right]
\cosh \left\{ \frac{\beta}{2} \sqrt{\left[J \left(\sigma_{i}^z + \sigma_{i+1}^z\right) + H_2^z \right]^2 + \left(H_2^x\right)^2} \right\}. \label{tm}
\end{eqnarray}
The effective Boltzmann's factor~(\ref{tm}) apparently depends, after tracing out the spin degrees of freedom of the $i$th Heisenberg spin,
only upon its two nearest-neighbour Ising spins $\sigma_{i}$ and $\sigma_{i+1}$. Thus, the expression~(\ref{tm}) can alternatively be viewed as the effective
two-by-two transfer matrix
\begin{eqnarray}
\mbox{T}\,\left(\sigma_{i}^z, \sigma_{i+1}^z\right) = \left( \!
\begin{array}{cc}
\mbox{T}\left(+\frac{1}{2}, +\frac{1}{2}\right)  & \mbox{T}\left(+\frac{1}{2}, -\frac{1}{2}\right) \\
\mbox{T}\left(-\frac{1}{2}, +\frac{1}{2}\right)  & \mbox{T}\left(-\frac{1}{2}, -\frac{1}{2}\right) \\
\end{array}
\! \right)
= \left( \!
\begin{array}{cc}
{T}_{+} & {T}_0 \\
{T}_0  & {T}_{-} \\
\end{array}
\! \right)
\label{tmm}
\end{eqnarray}
with three different matrix elements defined as
\begin{eqnarray}
T_{\pm} \!\!\!&\equiv&\!\!\! \mbox{T}\left(\pm\frac{1}{2},\pm\frac{1}{2}\right) =  2 \exp \left(\pm \frac{\beta}{2} H_1^z \right)
\cosh \left[ \frac{\beta}{2} \sqrt{\left(J \pm H_2^z \right)^2 + \left(H_2^x\right)^2} \right], \nonumber \\
T_0 \!\!\!&\equiv&\!\!\! \mbox{T}\left(\pm\frac{1}{2},\mp\frac{1}{2}\right) = 2 \cosh \left[ \frac{\beta}{2} \sqrt{\left(H_2^z\right)^2 + \left(H_2^x\right)^2} \right].
\label{tme}
\end{eqnarray}
Substituting the matrix~(\ref{tmm}) into the relation~(\ref{pf}) one may consequently employ the standard transfer-matrix approach
in order to obtain the exact result for the partition function
\begin{eqnarray}
{\cal Z} = \sum_{\{ \sigma_i \}} \prod_{i=1}^{N} \mbox{T}\,(\sigma_{i}^z, \sigma_{i+1}^z) = \mbox{Tr} \, \mbox{T}^{N} = \lambda_1^N + \lambda_2^N,
\label{pfr}
\end{eqnarray}
which is written in terms of two respective eigenvalues of the transfer matrix~(\ref{tmm})
\begin{eqnarray}
\lambda_{1,2} = \frac{1}{2} \left[T_{+} + T_{-} \pm \sqrt{\left(T_{+} - T_{-}\right)^2 + 4 T_0^2} \right].
\label{etm}
\end{eqnarray}

Now, let us proceed to the calculation of the most important quantities, which are relevant for our subsequent analysis of the magnetization process.
The Gibbs free energy can easily be calculated from the exact expression~(\ref{pfr}) for the partition function.
In the thermodynamic limit $N \to \infty$, one  obtains the following precise analytical result for the free energy per elementary unit
\begin{eqnarray}
f = - k_{\rm B} T \lim_{N \to \infty} \frac{1}{N} \ln {\cal Z} = k_{\rm B} T \ln 2
    - k_{\rm B} T \ln \left[T_{+} + T_{-} + \sqrt{(T_{+} - T_{-})^2 + 4 T_0^2} \right].
\label{free}
\end{eqnarray}
Subsequently, one may readily calculate the sublattice
magnetizations of the Ising and Heisenberg spins by differentiating the free energy (\ref{free}) with respect to the appropriate effective fields.
The sublattice magnetization of the Ising spins in two mutually orthogonal directions of the external magnetic field oriented either
perpendicular or parallel with respect to the easy axis read
\begin{eqnarray}
m_1^x \equiv g_1^x \mu_{\rm B} \langle \sigma_i^x \rangle = 0, \qquad
m_1^z \equiv g_1^z \mu_{\rm B} \langle \sigma_i^z \rangle = \frac{g_1^z \mu_{\rm B}}{2} \frac{T_{+} - T_{-}}{\sqrt{(T_{+} - T_{-})^2 + 4T_0^2}}\,.
\label{mi}
\end{eqnarray}
Similarly, the sublattice magnetization of the Heisenberg spins in two aforementioned orthogonal directions of the external
magnetic field can be calculated from the following unique formula
\begin{eqnarray}
m_2^{\alpha} \equiv g_2^{\alpha} \mu_{\rm B} \langle S^{\alpha}_i \rangle = \frac{g_2^{\alpha} \mu_{\rm B}}{2}
\frac{\left(T_{+} - T_{-}\right)\left(U_{+}^{\alpha} - U_{-}^{\alpha}\right) +  4 T_0 U_0^{\alpha} + \left(U_{+}^{\alpha} + U_{-}^{\alpha}\right)\sqrt{(T_{+} - T_{-})^2 + 4 T_0^2}}
     {(T_{+} - T_{-})^2 + 4 T_0^2 + (T_{+} + T_{-})\sqrt{(T_{+} - T_{-})^2 + 4 T_0^2}} \, ,
\label{mh}
\end{eqnarray}
which is valid for both spatial directions $\alpha = x$ and $z$ if the coefficients $U_{\pm}^{\alpha}$ and $U_0^{\alpha}$ are defined as
\begin{eqnarray}
U_{\pm}^x \!\!\!&=&\!\!\! \frac{H_2^x}{\sqrt{\left(J \pm H_2^z\right)^2 + \left(H_2^x\right)^2}}
                          2 \exp \left(\pm \frac{\beta H_1^z}{2} \right) \sinh \left[ \frac{\beta}{2} \sqrt{\left(J \pm H_2^z\right)^2 + \left(H_2^x\right)^2} \right], \nonumber \\
U_0^x \!\!\!&=&\!\!\! \frac{H_2^x}{\sqrt{\left(H_2^z\right)^2 + \left(H_2^x\right)^2}} 2 \sinh \left[ \frac{\beta}{2} \sqrt{\left(H_2^z\right)^2 + \left(H_2^x\right)^2} \right],
\label{coef1}
\end{eqnarray}
and
\begin{eqnarray}
U_{\pm}^z \!\!\!&=&\!\!\! \pm \frac{\left(J \pm H_2^z\right)}{\sqrt{\left(J \pm H_2^z\right)^2 + \left(H_2^x\right)^2}}
                          2  \exp \left(\pm \frac{\beta H_1^z}{2} \right) \sinh \left[ \frac{\beta}{2} \sqrt{\left(J \pm H_2^z\right)^2 + \left(H_2^x\right)^2} \right], \nonumber \\
U_0^z \!\!\!&=&\!\!\! \frac{H_2^z}{\sqrt{\left(H_2^z\right)^2 + \left(H_2^x\right)^2}} 2 \sinh \left[ \frac{\beta}{2} \sqrt{\left(H_2^z\right)^2 + \left(H_2^x\right)^2} \right].
\label{coef2}
\end{eqnarray}

It should be pointed out that the exact analytical formula for the free energy~(\ref{free}) permits a straightforward derivation of the magnetization formula for the most general case of arbitrarily oriented external magnetic field as well. The final magnetization formula in the external field of an arbitrary spatial direction can be conveniently expressed in terms of the formerly derived sublattice magnetizations~(\ref{mi}) and~(\ref{mh}) of the Ising and Heisenberg spins
\begin{eqnarray}
m (\theta) =  \left(m_1^z + m_2^z\right) \cos \theta + m_2^x \sin \theta.
\label{scs}
\end{eqnarray}
It is worth noting that the magnetization formula~(\ref{scs}) might be of fundamental importance for the analysis of the angular dependence of the magnetization process in a single-crystal sample related to the considered spin-chain model. On the other hand, the magnetization of a powder sample pertinent to the investigated spin-chain model
can also be formally obtained by integrating the magnetization formula~(\ref{scs}) over one hemisphere yielding
\begin{eqnarray}
m_{\rm p} = \int_{0}^{\frac{\pi}{2}} \! \! m (\theta) \sin \theta \, {\rm d} \theta,
\label{pcs}
\end{eqnarray}
but the involved integral precludes derivation of the closed-form analytical expression due to too complicated functions involved in the sublattice magnetizations~(\ref{mi}) and~(\ref{mh}) of the Ising and Heisenberg spins, respectively. Of course, the integral entering the relation~(\ref{pcs}) for the magnetization of powder samples
can be evaluated numerically and hence, it may be of practical importance for an investigation of the magnetization process in the related polycrystalline systems.

\section{Results and discussion}
\label{result}

Let us proceed to a discussion of the most interesting findings acquired for the ferrimagnetic spin-$\frac{1}{2}$ chain of alternating Ising and Heisenberg spins coupled through the antiferromagnetic nearest-neigh\-bour interaction $J=-|J|<0$. First, we will present a comprehensive survey of theoretical results with the aim to shed light on typical magnetization features of the proposed spin-chain model and then, high-field magnetization data of one specific polymeric coordination compound will be clarified within the framework of the model under investigation.

\subsection{Survey of theoretical results}

Let us begin with the analysis of the ground state. The diagonal form of the site Hamiltonian~(\ref{hamid}) allows one to get the lowest-energy eigenstate of the investigated spin alternating chain
\begin{eqnarray}
| \mbox{FRI} \rangle = \prod_{i=1}^{N} \Big| \sigma_i^{z} = \frac{1}{2} \Big \rangle  \left( c_{-} \, \Big| S_i^{z} = -\frac{1}{2} \Big\rangle + c_{+} \, \Big| S_i^{z} = \frac{1}{2} \Big \rangle \right),
\label{gs}
\end{eqnarray}
which indicates the \textit{quantum ferrimagnetic ordering} with a perfect alignment of the Ising spins towards their easy axis and, respectively,
the quantum superposition of two spin states of each Heisenberg spin that basically depends on a mutual interplay between the exchange constant, Land\'e $g$-factor,
size and spatial orientation of the external field via the occurrence probabilities
\begin{eqnarray}
c_{\pm}^2 = \frac{1}{2} \left[1 \mp \frac{|J| - g_2^z \mu_{\rm B} B \cos \theta}{\sqrt{\left(|J| - g_2^z \mu_{\rm B} B \cos \theta \right)^2
+ \left(g_2^x \mu_{\rm B} B \sin \theta \right)^2}}\right].
\label{pa}
\end{eqnarray}
The perfect alignment of the Ising spins is also confirmed by the following zero-temperature values of the relevant sublattice magnetization
in two conspicuous orthogonal directions (parallel and perpendicular) with respect to the easy axis
\begin{eqnarray}
m_1^z = \frac{1}{2} g_1^z \mu_{\rm B} \, , \qquad  m_1^x = 0.
\label{gsma}
\end{eqnarray}
Contrary to this, both spatial components of the sublattice magnetization of the Heisenberg spins are subject to the quantum reduction of magnetization on behalf
of local quantum fluctuations arising from the transverse field (i.e., perpendicular projection of the external magnetic field with respect to the easy axis of the Ising spins)
\begin{eqnarray}
m_2^z \!\!\!&=&\!\!\! - \frac{g_2^z \mu_{\rm B}}{2} \left[\frac{|J| - g_2^z \mu_{\rm B} B \cos \theta}{\sqrt{\left(|J| - g_2^z \mu_{\rm B} B \cos \theta \right)^2
+ \left(g_2^x \mu_{\rm B} B \sin \theta \right)^2}}\right],\nonumber  \\
m_2^x \!\!\!&=&\!\!\! \phantom{-}\frac{g_2^x \mu_{\rm B}}{2} \left[\frac{g_2^x \mu_{\rm B} B \sin \theta}{\sqrt{\left(|J| - g_2^z \mu_{\rm B} B \cos \theta \right)^2
+ \left(g_2^x \mu_{\rm B} B \sin \theta \right)^2}}\right].
\label{gsmb}
\end{eqnarray}
It is quite obvious from~(\ref{gsma}) and~(\ref{gsmb}) [or alternatively from~(\ref{gs}) and~(\ref{pa})] that the quantum ferrimagnetic order changes to the classical ferrimagnetic order whenever the transverse projection of the external magnetic field vanishes. However, the total magnetization of the ferrimagnetic chain of alternating
Ising and Heisenberg spins is subject
to the quantum reduction of magnetization for any other spatial orientation of the external magnetic field
\begin{eqnarray}
m \!\!\!&=&\!\!\!\frac{\mu_{\rm B} \cos \theta}{2} \left[g_1^z  - g_2^z \frac{|J| - g_2^z \mu_{\rm B} B \cos \theta}{\sqrt{\left(|J| - g_2^z \mu_{\rm B} B \cos \theta \right)^2
+ \left(g_2^x \mu_{\rm B} B \sin \theta \right)^2}}\right]  \nonumber\\
\!\!\!&+&\!\!\! \frac{g_2^x \mu_{\rm B} \sin \theta}{2} \left[\frac{g_2^x \mu_{\rm B} B \sin \theta}{\sqrt{\left(|J| - g_2^z \mu_{\rm B} B \cos \theta \right)^2 + \left(g_2^x \mu_{\rm B} B \sin \theta \right)^2}}\right],
\label{gsmt}
\end{eqnarray}
which appears due to the quantum entanglement of two spin states of each Heisenberg spin arising from the transverse field. In accordance with this statement, the quantum reduction of total magnetization becomes  greater, the higher the transverse component of the external field is.

\begin{figure}
\vspace{-0.5cm}
\begin{center}
\includegraphics[width=6.4cm]{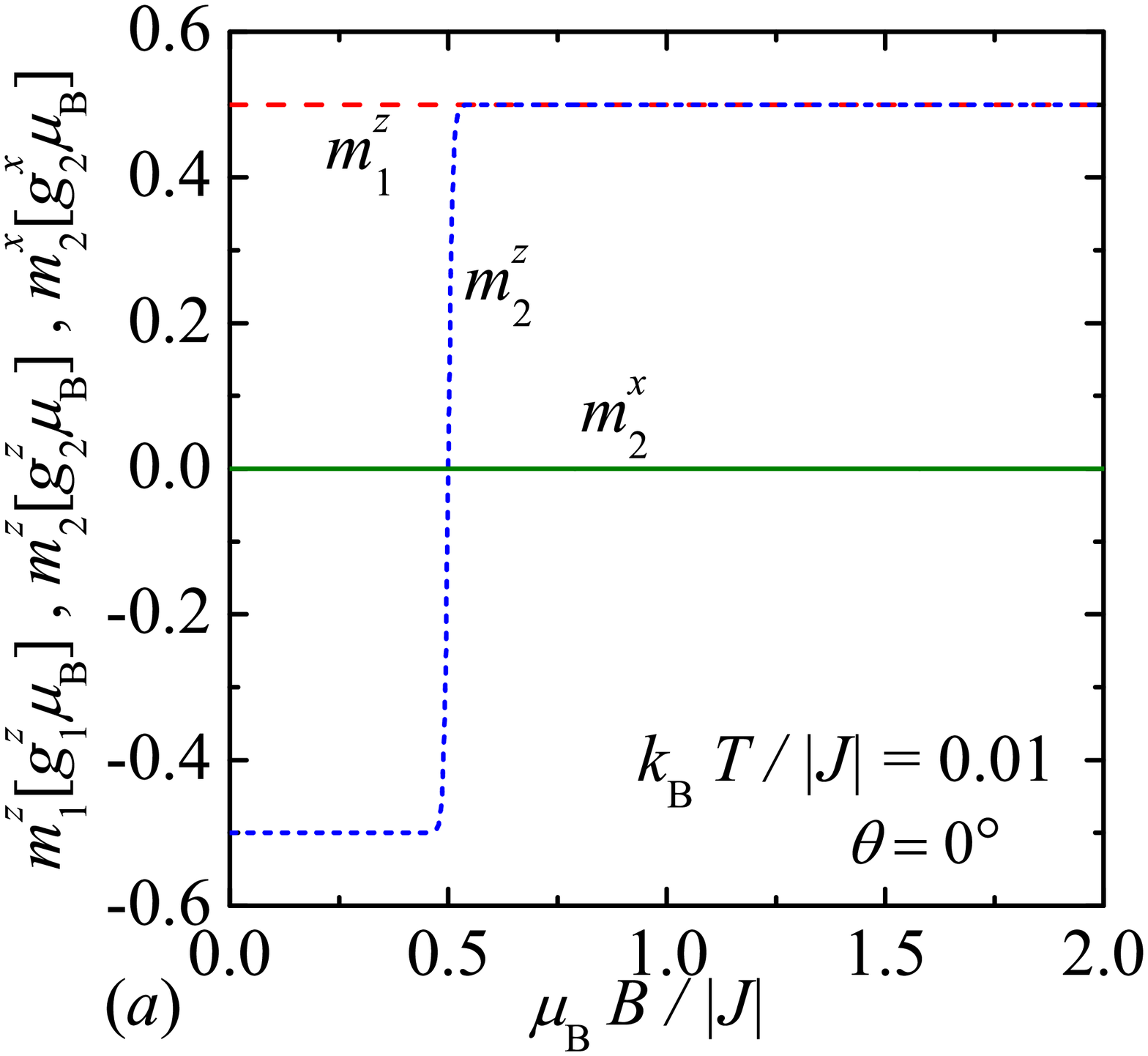}
\hspace{-2.35cm}
\includegraphics[width=6.4cm]{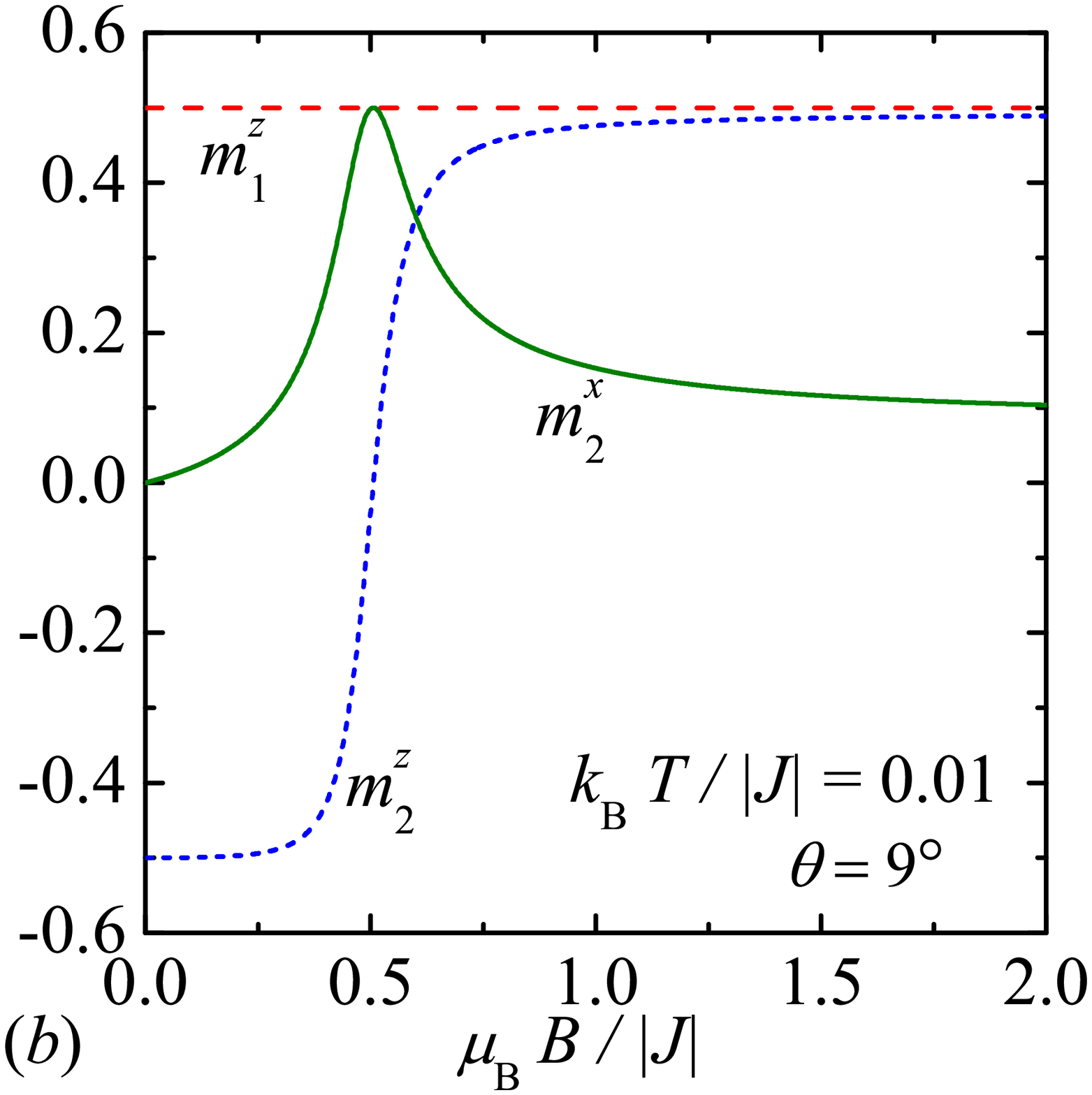}
\hspace{-2.35cm}
\includegraphics[width=6.4cm]{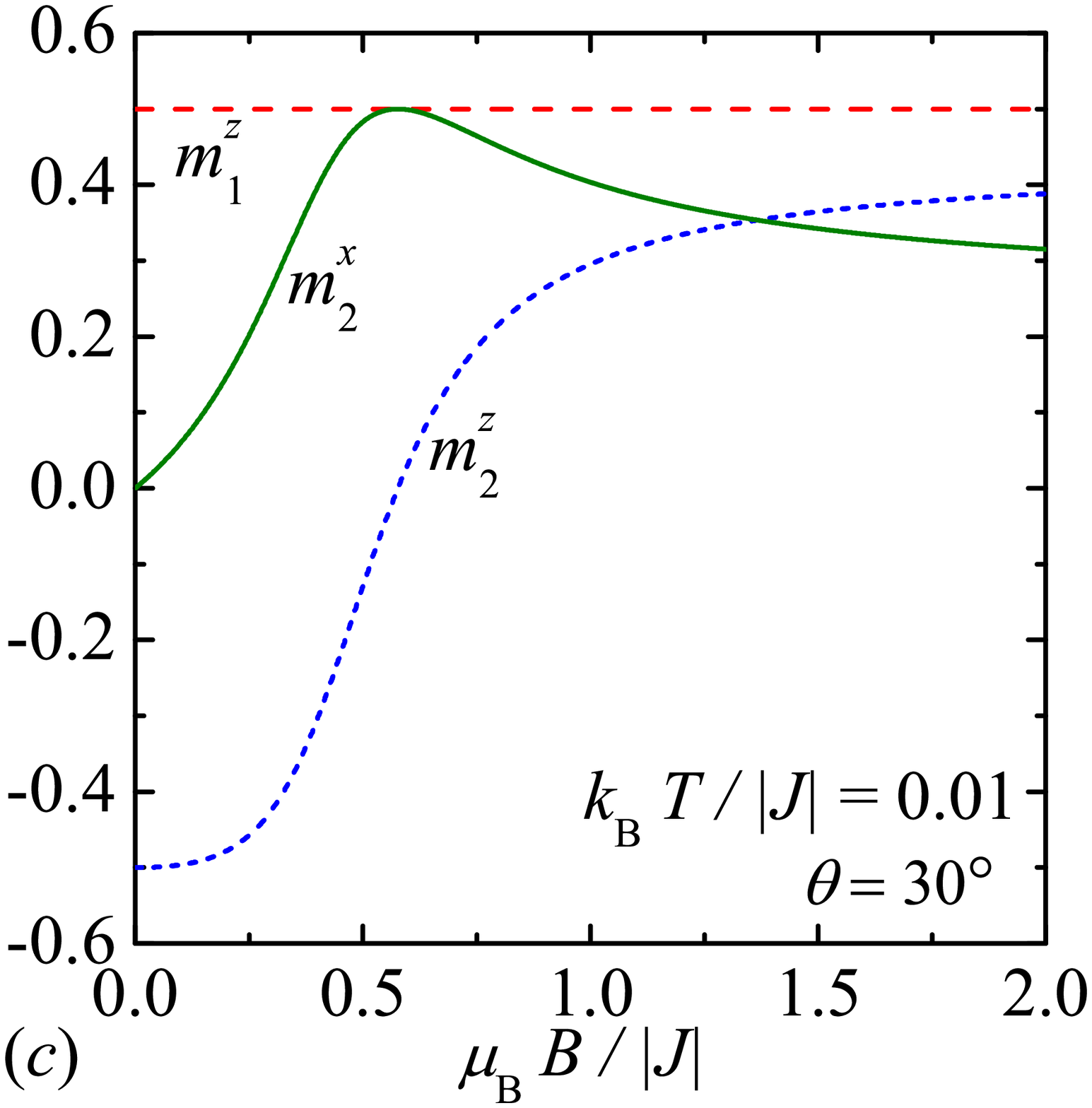}
\includegraphics[width=6.4cm]{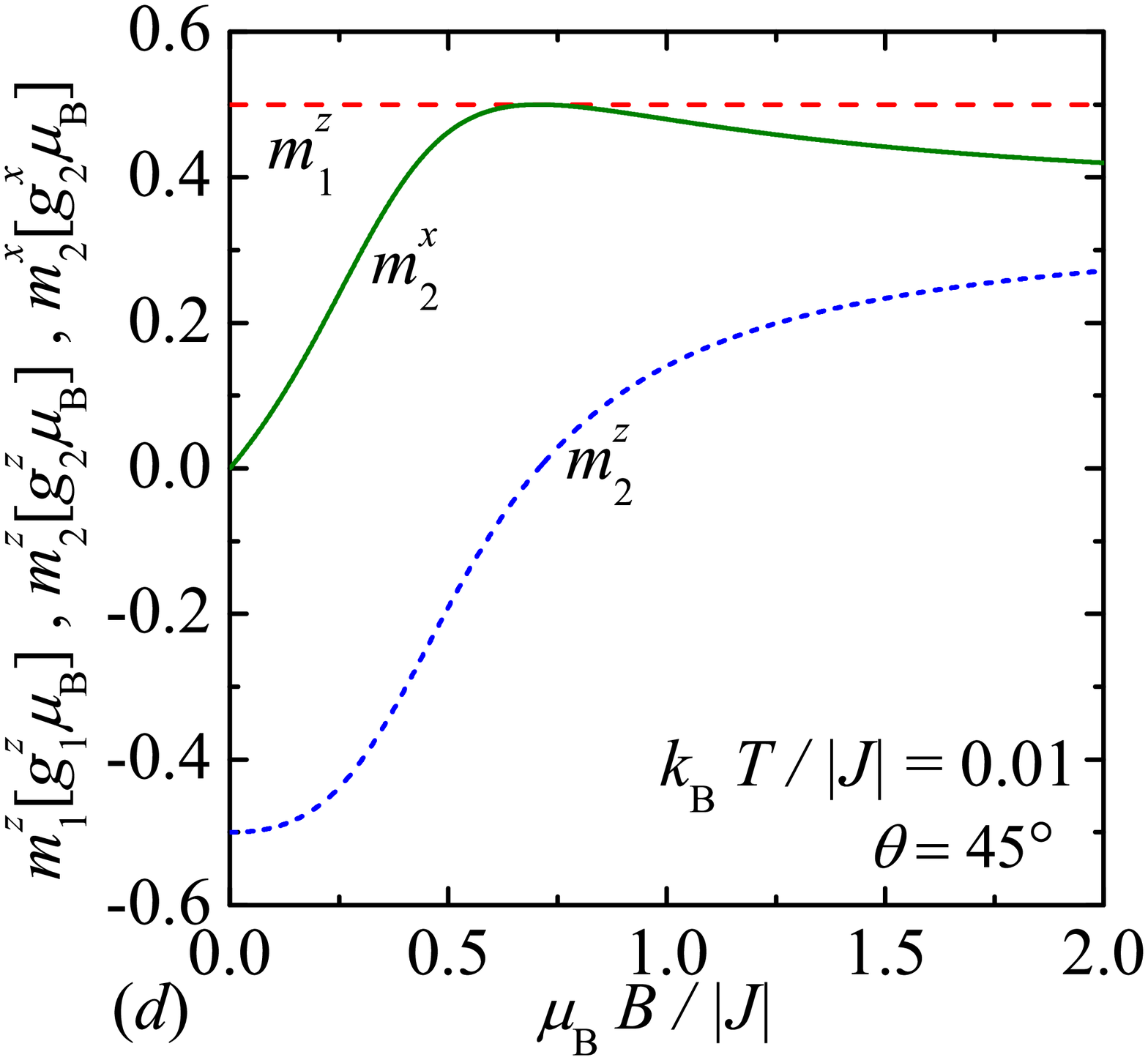}
\hspace{-2.35cm}
\includegraphics[width=6.4cm]{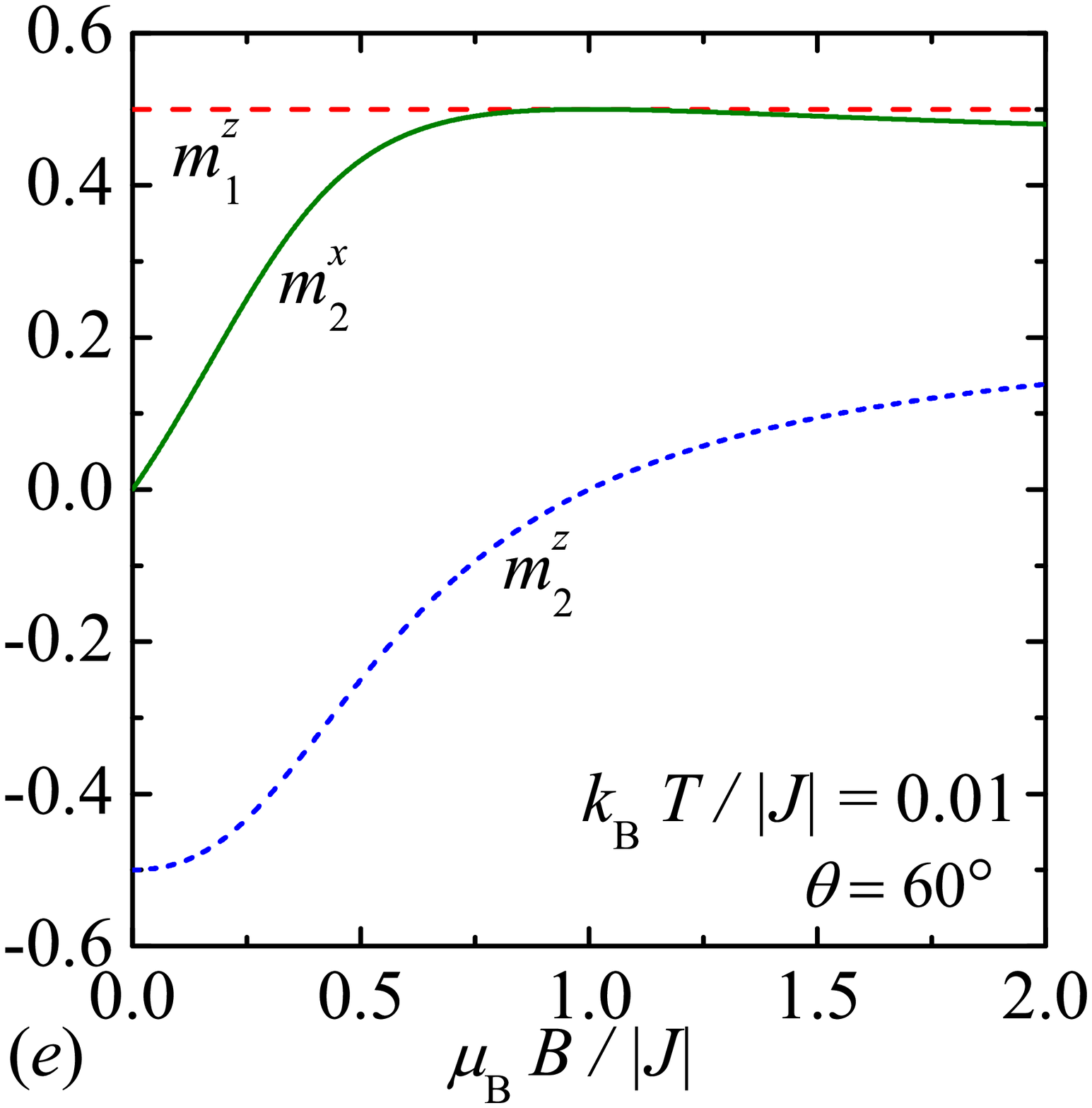}
\hspace{-2.35cm}
\includegraphics[width=6.4cm]{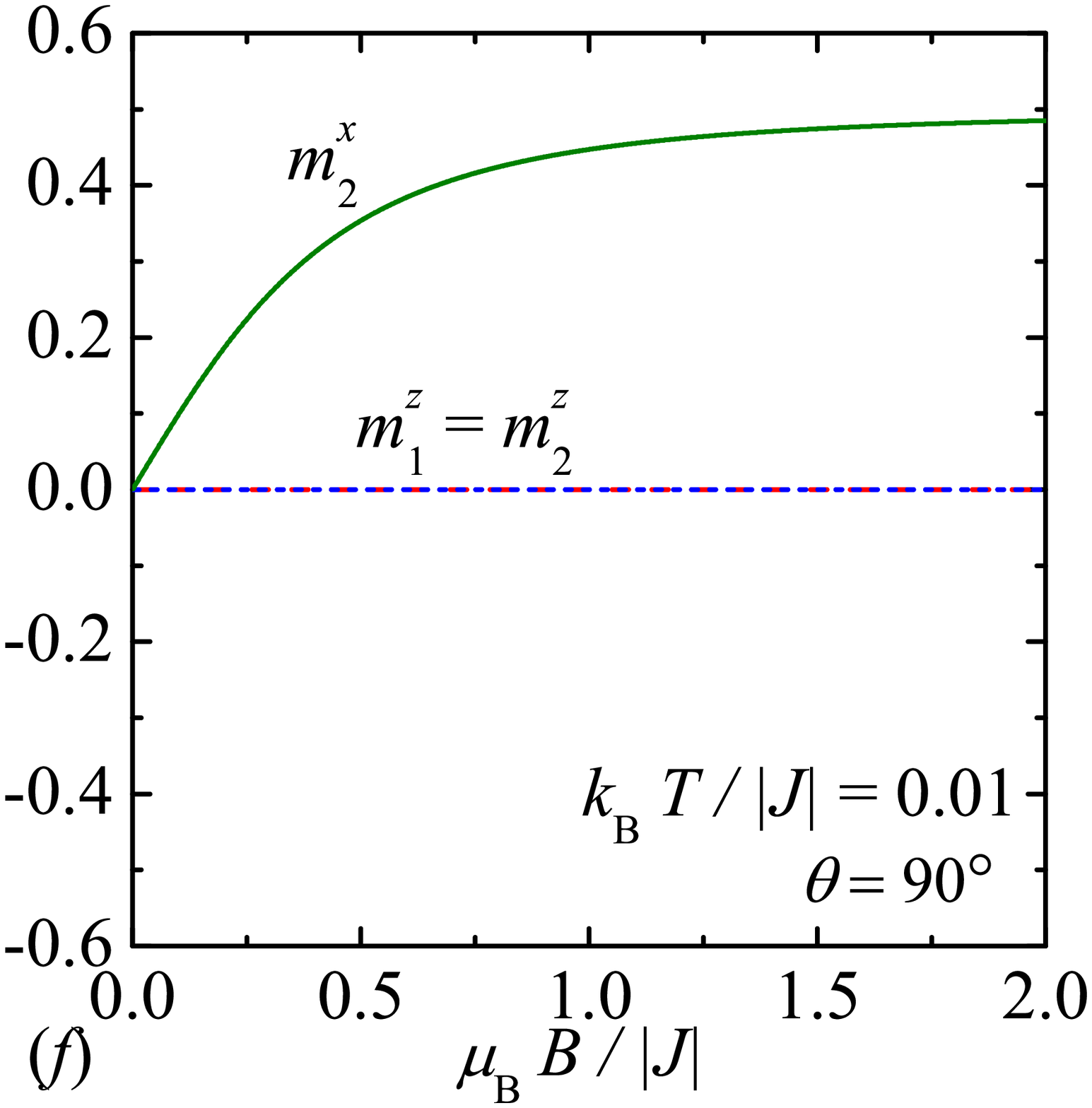}
\end{center}
\vspace{-0.5cm}
\caption{(Color online) Longitudinal and transverse projections of the sublattice magnetization of the Ising ($m_1^z$, $m_1^x=0$) and Heisenberg ($m_2^z$, $m_2^x$) spins as a function of the magnetic field at low enough temperature $k_{\rm B} T/ |J| = 0.01$, one particular choice of $g$-factors $g_1^z = 6$, $g_1^x = 0$, $g_2^x = g_2^z = 2$, and several spatial orientations of the external magnetic field.}
\label{fig1}
\end{figure}

\begin{figure}
\vspace{-0.5cm}
\begin{center}
\includegraphics[width=7.5cm]{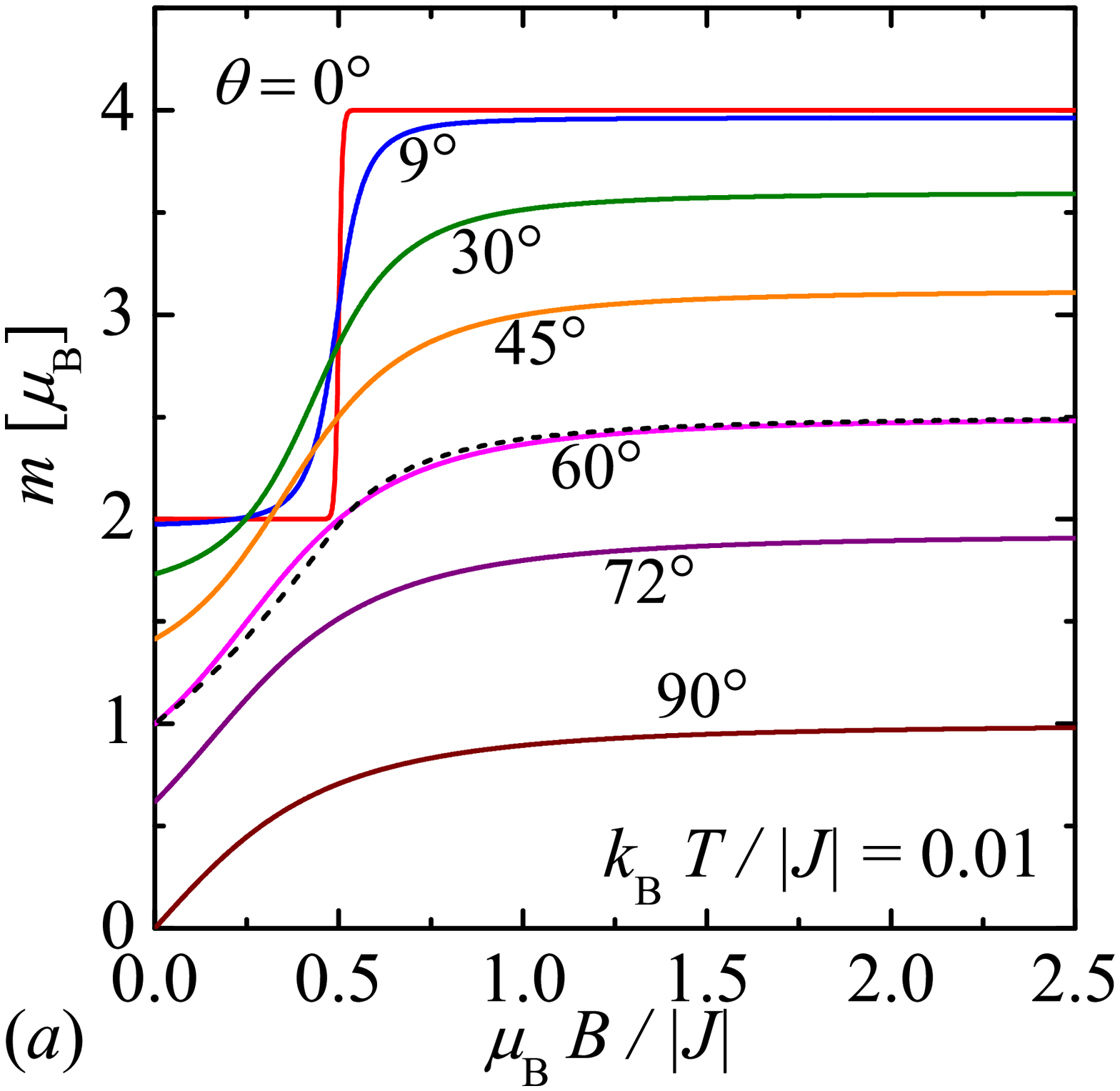}
\hspace{-2.0cm}
\includegraphics[width=7.5cm]{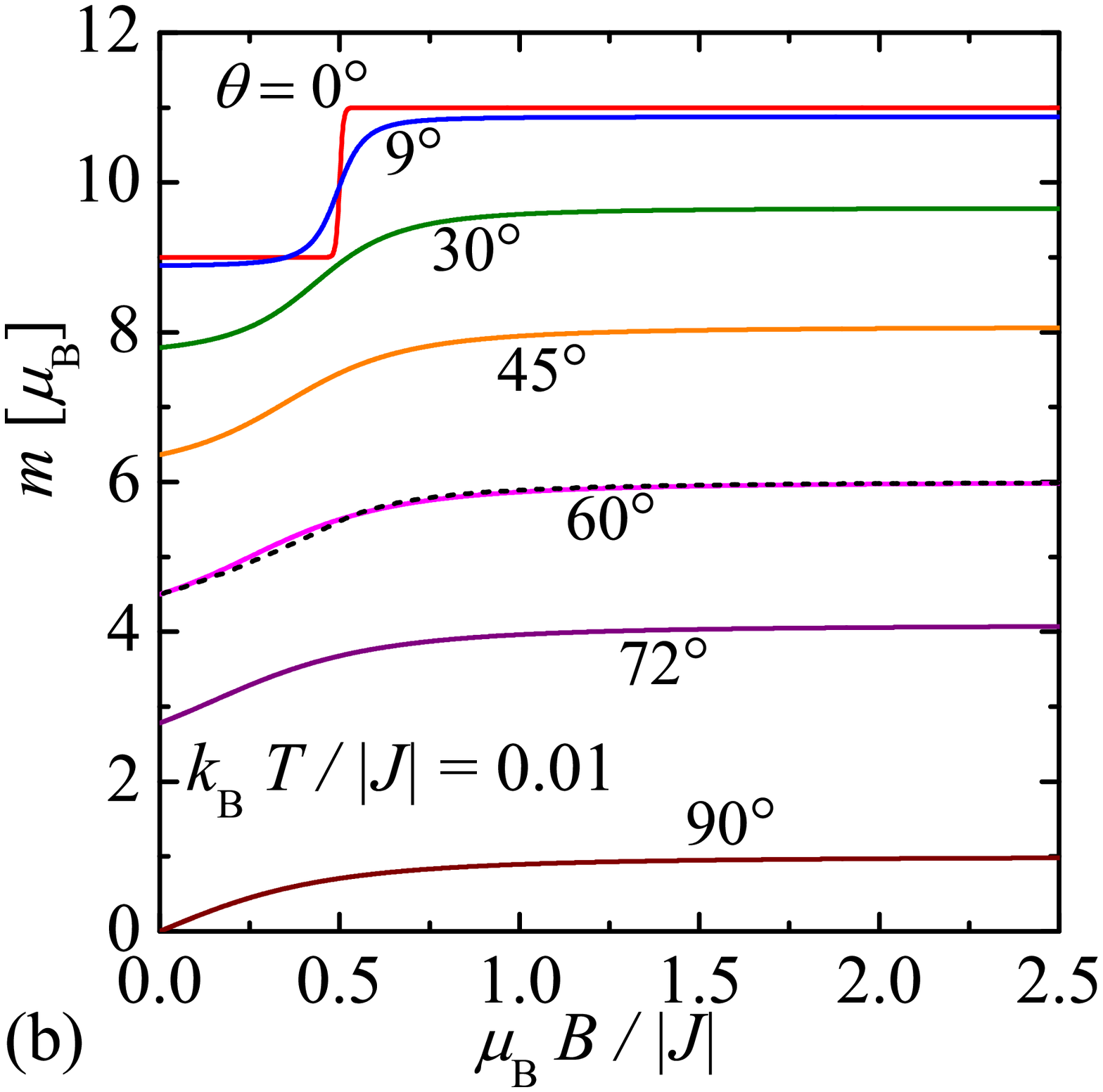}
\end{center}
\vspace{-0.6cm}
\caption{(Color online) Low-temperature dependence of the total magnetization as a function of the magnetic field for several spatial orientations of the external field and two different sets of $g$-factors: (a) $g_1^z = 6$, $g_1^x = 0$, $g_2^x = g_2^z = 2$; (b) $g_1^z = 20$, $g_1^x = 0$, $g_2^x = g_2^z = 2$. Broken lines show magnetization curves for the powder samples, which were obtained by the numerical solution of~(\ref{pcs}).}
\label{fig2}
\end{figure}

Now, let us turn our attention to a discussion of the low-temperature magnetization process serving in evidence of the aforedescribed ground-state features. For simplicity, we will further assume that the Heisenberg spins have the completely isotropic Land\'e factor $g_2^x = g_2^y = g_2^z = 2$ in contrast to the highly anisotropic Land\'e factor of Ising spins $g_1^z \gg 2$ and $g_1^x=g_1^y=0$. To provide an in-depth understanding of the magnetization process, figure~\ref{fig1} illustrates typical field dependences for two orthogonal projections of the sublattice magnetization of the Ising and Heisenberg spins under different spatial orientation of the applied magnetic field. If the external field is applied along the easy axis of the Ising spins, one observes a steep field-induced increase in the longitudinal sublattice magnetization of the Heisenberg spins $m_2^z$ at a critical field due to an abrupt reversal of all Heisenberg spins towards the external-field direction (at low fields, the Heisenberg spins are aligned antiparallel with respect to the external field, because they possess lower $g$-factor compared to that of the Ising spins). This classical mechanism for the formation of an intermediate magnetization plateau will consequently lead to a sharp stepwise profile in the total magnetization versus the external-field dependence (see figure~\ref{fig2} and the subsequent discussion). If the external field is tilted from the easy-axis direction, the relevant behaviour of the Heisenberg spins becomes much more intricate owing to a mutual competition between two different spatial directions given by the easy axis of the Ising spins and the spatial orientation of the applied magnetic field. As a matter of fact, the longitudinal component of the sublattice magnetization of the Heisenberg spins $m_2^z$ apparently exhibits a smoother variation if the external field is gradually tilted from the easy axis and the more gentle field-induced reversal of the Heisenberg spins (given by the change of sign of $m_2^z$) simultaneously appears at the higher critical field
\begin{eqnarray}
\frac{\mu_{\rm B} B_{\rm c}}{|J|} = \frac{1}{g_2^z \cos \theta} \, .
\label{hcrit}
\end{eqnarray}
It should be mentioned that the transverse projection of the sublattice magnetization of the Heisenberg spins $m_2^x$ becomes non-zero for this more general case
and $m_2^x$ exhibits a striking non-monotonous dependence with a more or less sharper global maximum located at the critical field~(\ref{hcrit}). The particular case of the external magnetic field oriented perpendicular with respect to the easy axis deserves a special mention. Under this specific constraint, the Ising spins become completely uncorrelated (disordered) and the Heisenberg spins gradually tend to align towards the external-field direction upon the strengthening of the transverse field.

In figure~\ref{fig2}, the total magnetization is plotted against the magnetic field for various spatial orientations of the external field and two different values of the Land\'e $g$-factor of the Ising spins. It is worth noting that the profile of the displayed magnetization curves can readily be understood from the relevant low-temperature dependences of the sublattice magnetizations (cf. figure~\ref{fig2} with figure~\ref{fig1}). In agreement with our expectations, the classical mechanism for the formation of the intermediate magnetization plateau becomes quite evident under the particular orientation of the external field directed along the easy axis of the Ising spins. Under this specific condition, one actually detects a sharp stepwise magnetization curve with a marked intermediate plateau at the fractional value $(g_1^z-g_2^z) \mu_{\rm B}/2$ per elementary unit, which implies the existence of the classical ferrimagnetic order due to an unequal magnetic moment of the nearest-neighbour spin-$\frac{1}{2}$ atoms
differing in their $g$-factors. It can be clearly seen from figure~\ref{fig2} that the intermediate magnetization plateau gradually shrinks and the relevant dependence of the total magnetization becomes smoother as the external field deviates from the easy-axis direction of the Ising spins. The observed breakdown of intermediate magnetization plateau, the gradual smoothing of the magnetization curve as well as the overall quantum reduction of the total magnetization can all be attributed to local quantum fluctuations, which arise from the transverse component of the external magnetic field acting on the Heisenberg spins. For comparison, the low-temperature magnetization curve of a polycrystalline system is also depicted in figure~\ref{fig2} by a broken line. Interestingly, the powder averaging through the formula~(\ref{pcs}) yields the magnetization curve of a polycrystalline system, which quite closely follows the magnetization curve of a single-crystal system for the particular orientation of the external magnetic field deviating by the angle $\theta = 60^{\circ}$ from the easy-axis direction.

\subsection{High-field magnetization of DyCu}

\begin{figure}[!t]
\begin{center}
\includegraphics[width=12cm]{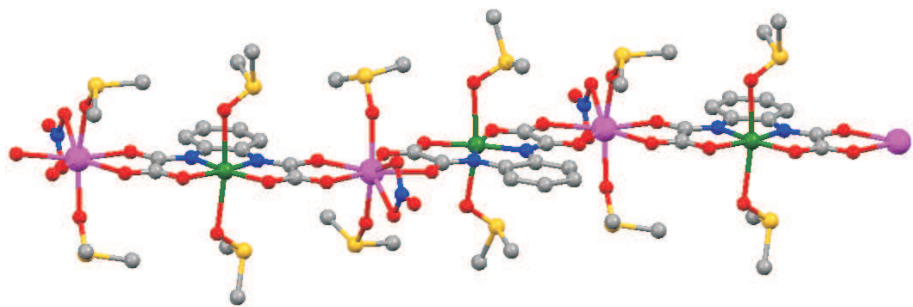}
\end{center}
\vspace*{-0.3cm}
\caption{(Color online) The visualization of LnCu polymeric chain in bimetallic coor\-dina\-tion com\-pounds Ln(NO$_3$)(DMSO)$_2$Cu(opba)(DMSO)$_2$ (Ln $=$ Gd--Er) by adopting
the crystallographic data from reference~\cite{cal08} deposited at The Cambridge Crystallographic Data Centre. Coloring scheme for the atom labelling:
Ln (magenta), Cu (green), O (red), N (blue), C (grey), S (yellow).}
\label{fig3}
\end{figure}

Our theoretical findings for the magnetization process will be now confronted with high-field magnetization data of $3d$-$4f$ bimetallic coordination polymer Dy(NO$_3$)(DMSO)$_2$Cu(opba)(DMSO)$_2$ (DMSO $=$ dimethylsulfoxide, opba $=$ orthophenylenebisoxamato) to be further abbreviated as DyCu. The polycrystalline
sample of DyCu was prepared according to the procedure reported previously by Calvez and co-workers~\cite{cal08}. Even though our efforts aimed at preparing a single-crystal sample suitable for a crystal-structure characterization and magnetization measurements was not successful, the elemental analysis has provided a strong
support that the prepared polycrystalline sample should consist of bimetallic polymeric chains representing a structural analogue of Ln(NO$_3$)(DMSO)$_2$Cu(opba)(DMSO)$_2$ (Ln $=$ Gd--Er), which is visualized in figure~\ref{fig3} by making use of the crystallographic data reported by Calvez {et al}.~\cite{cal08}. Bearing this in mind, it could be expected that the bis-chelating bridging ligand opba might mediate a relatively strong superexchange coupling between the nearest-neighbour Dy$^{3+}$ and Cu$^{2+}$ magnetic ions, which should consequently form the bimetallic polymeric chain running along the crystallographic $c$-axis.

First, let us make a few comments on the constituent magnetic ions of DyCu. It is quite well established that the magnetic behaviour of octahedrally coordinated Cu$^{2+}$ ions
can be quite faithfully represented by the notion of Heisenberg spins with a more or less isotropic Land\'e factor $g_2^x \approx g_2^y \approx g_2^z \gtrsim 2$~\cite{jon74}, while the eight-coordinated Dy$^{3+}$ ions often obey more subtle requirements of the Ising spins necessitating a highly anisotropic $g$-factor $g_1^z \gg 2$ and $g_1^x \approx g_1^y \approx 0$~\cite{heu10,wol00,jon74,jen91}. In fact, Dy$^{3+}$ ion represents Kramers ion with the ground-state multiplet $^{6}$H$_{15/2}$, which usually undergoes a rather strong crystal-field splitting into eight well-separated Kramers doublets~\cite{wol00,jon74}. In this respect, the magnetic behaviour of Dy$^{3+}$ ion can be often interpreted at low enough temperatures in terms of the Ising spin with the effective Land\'e factor $g_1^z \approx 20$ and $g_1^x \approx g_1^y \approx 0$~\cite{heu10,wol00,jon74,jen91}. With all this in mind, the coordination polymer DyCu could provide a suitable experimental realization of the spin-$\frac{1}{2}$ chain of the alternating Ising and Heisenberg spins. However, one should also keep in mind that this simplification is reasonable only at sufficiently low temperatures compared with the energy gap between the lowest-energy and excited Kramers doublets to avoid any danger of over-interpretation inherent in this approximation.

\begin{figure}[!b]
\vspace{-0.3cm}
\begin{center}
\includegraphics[width=7.5cm]{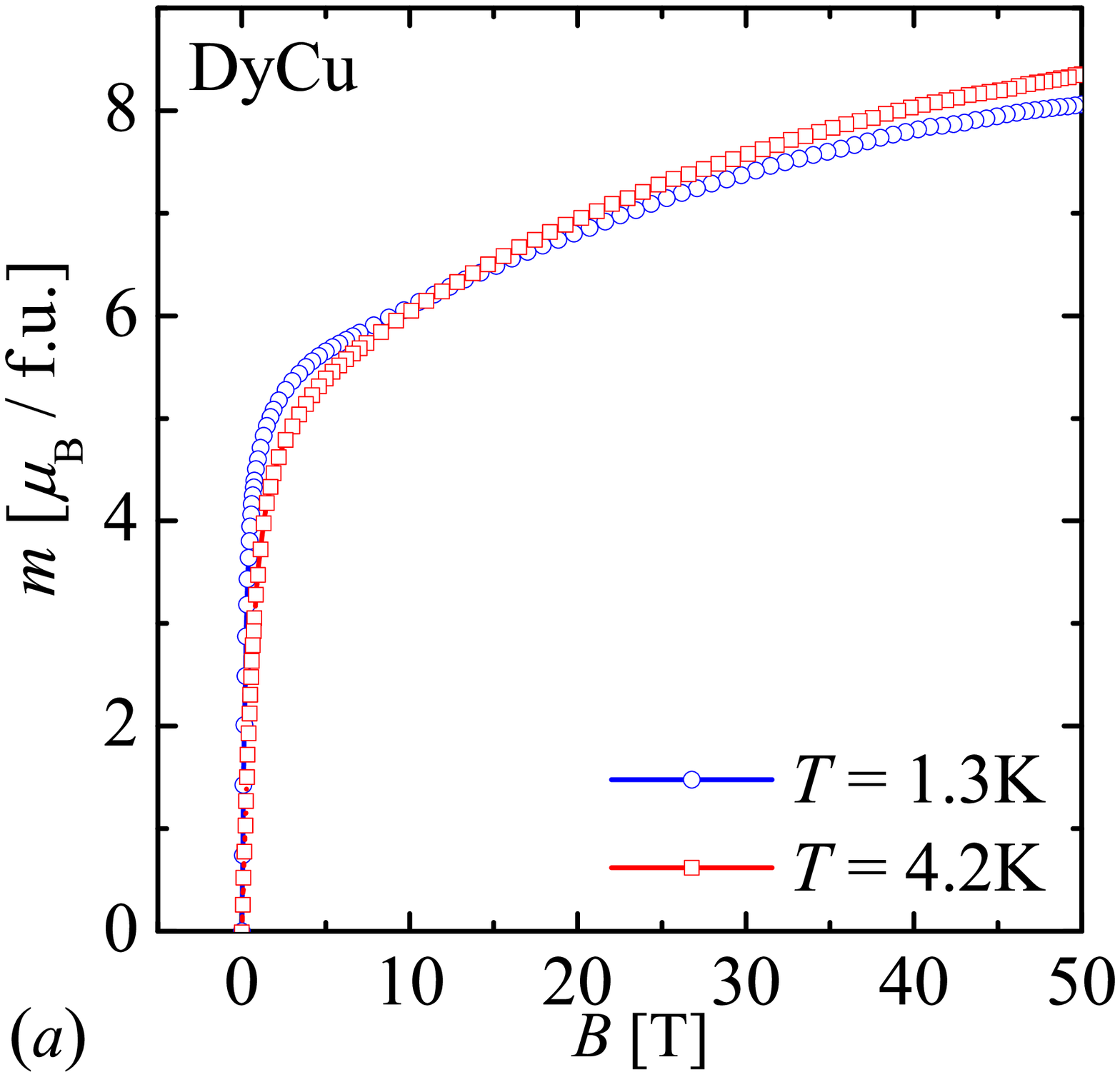}
\hspace{-2.0cm}
\includegraphics[width=7.5cm]{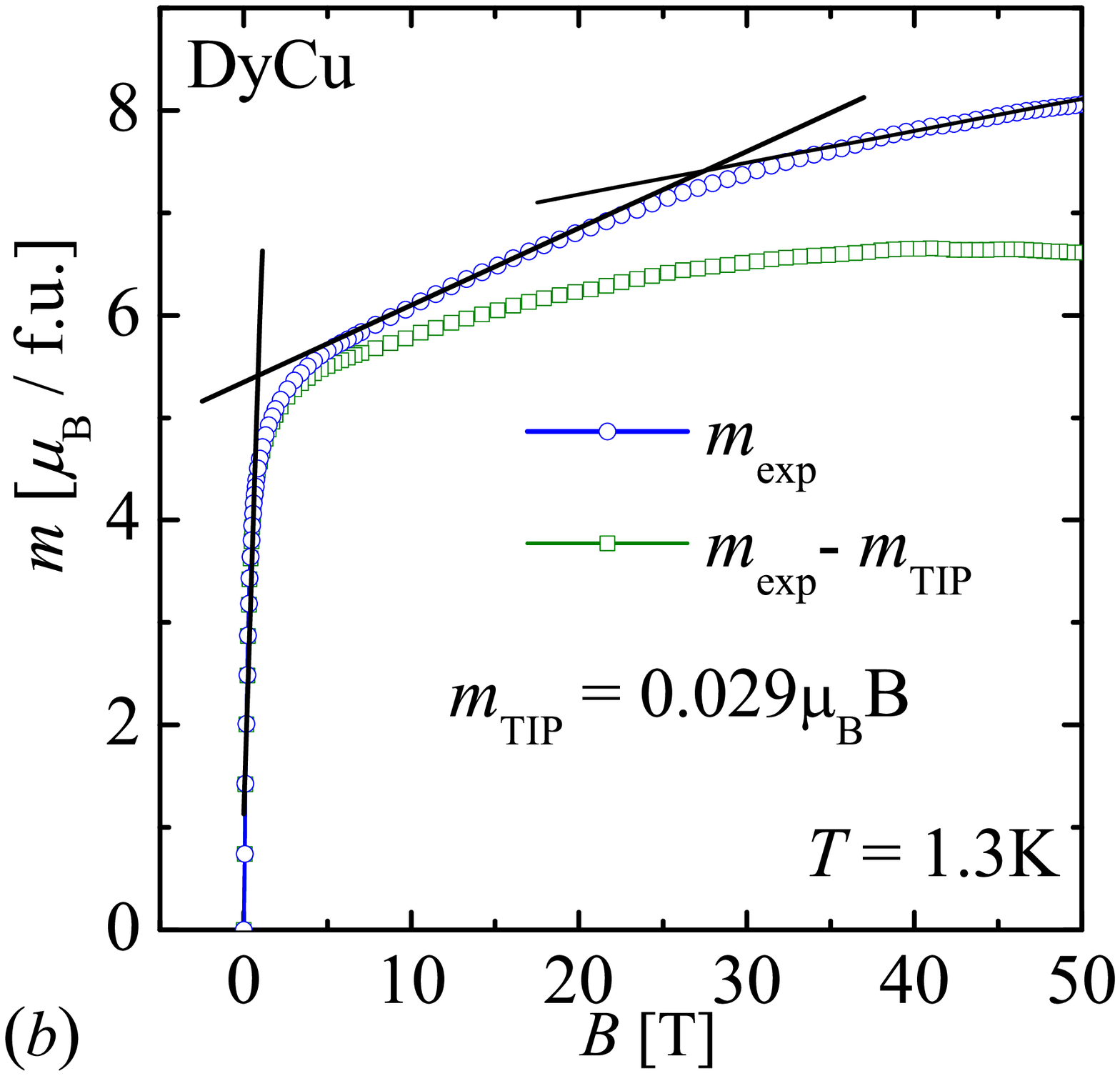}
\end{center}
\vspace{-0.3cm}
\caption{(Color online) (a) High-field magnetization data recorded for the powder sample of DyCu at two sufficiently low temperatures 1.3~K and 4.2~K;
(b) High-field magnetization curve of DyCu at the lowest measured temperature 1.3~K before and after subtracting the paramagnetic
Van Vleck contribution estimated from the quasi-linear dependence in the high-field range $36\div52$~T.}
\label{fig4}
\end{figure}

Figure \ref{fig4}~(a) shows the high-field magnetization curve of the powder sample of DyCu measured at two sufficiently low temperatures 1.3~K and 4.2~K (see reference~\cite{han12} for more details on the set-up of magnetization measurements). The displayed magnetization curves exhibit a remarkable crossing around 12~T, whereas the magnetization data recorded at a higher temperature 4.2~K become greater than the ones recorded at a lower temperature 1.3~K above this crossing field. Apart from this fact, one may clearly recognize three characteristic regions in the relevant magnetization process as illustrated in figure~\ref{fig4}~(b) by thin solid lines serving as guides for eyes only. The magnetization initially shows a very rapid increase with the magnetic field in the low-field range $0\div 3$~T, then it exhibits a rather steep increase in the range of
moderate fields $5\div 32$~T, which is consecutively followed with a more steady quasi-linear dependence in the high-field range $34\div 52$~T. The steady quasi-linear dependence
observed in the high-field range implies a significant contribution of the temperature-independent (Van Vleck) paramagnetism~\cite{jen91}, which was evaluated to be
$m_{\rm TIP}=0.029 \mu_{\rm B}~\textrm{T}^{-1}$ per Dy$^{3+}$ ion from the linear fit of the high-field region $36\div52$~T and subsequently subtracted from the measured magnetization data. Even though this procedure might give only a rather rough estimate of $m_{\rm TIP}$, the obtained value is in a rather good accord
with typical values of $m_{\rm TIP}$ reported previously for other compounds involving Dy$^{3+}$ ion~\cite{heu10}.

\begin{figure}[!h]
\begin{center}
\includegraphics[width=7.5cm]{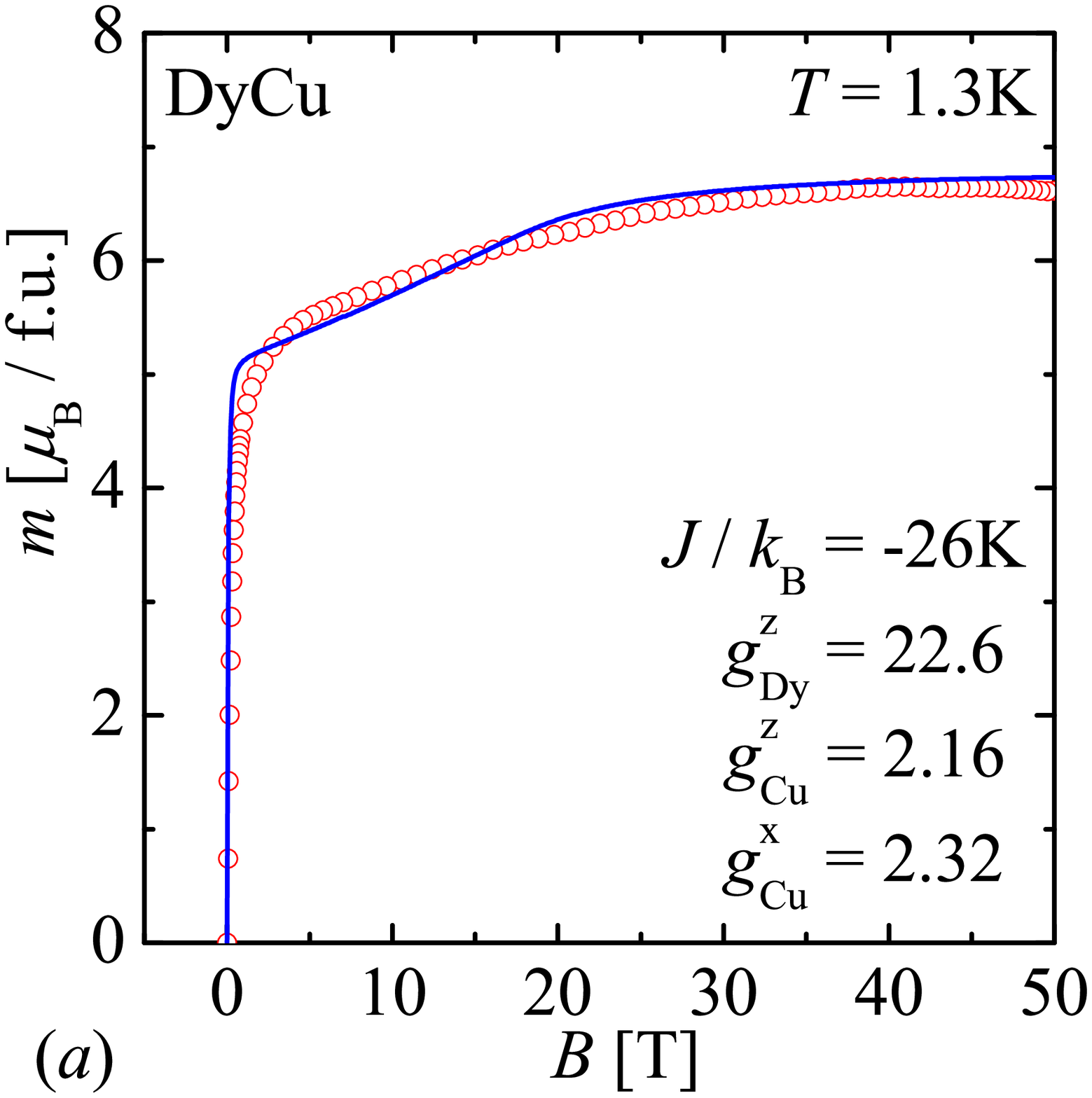}
\hspace{-2.0cm}
\includegraphics[width=7.5cm]{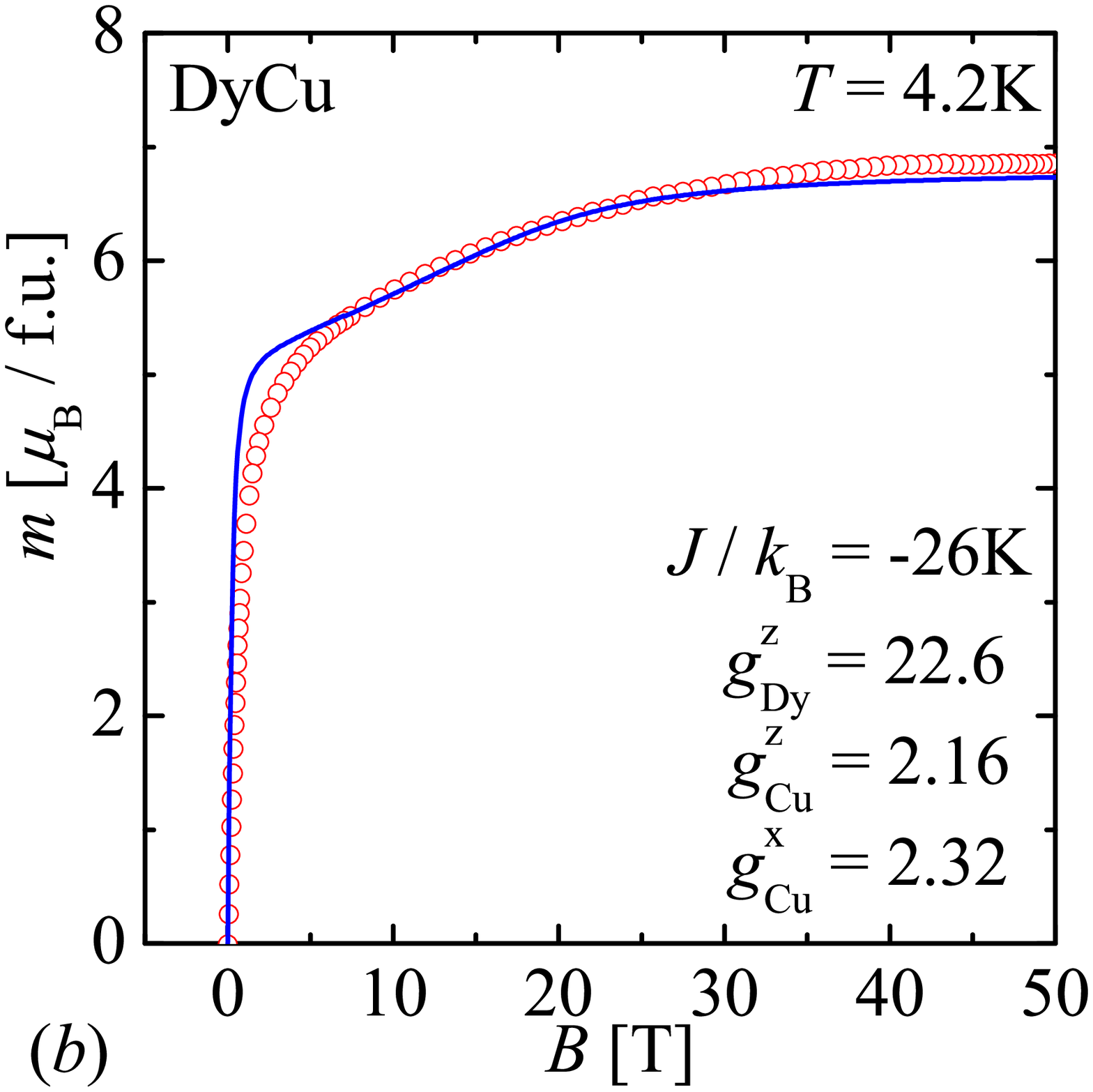}
\end{center}
\caption{(Color online) High-field magnetization data of the powder sample of DyCu at two different temperatures: (a) $T=1.3$~K; (b) $T=4.2$~K. Blue solid lines show
the best theoretical fit obtained using the formulas~(\ref{scs}) and~(\ref{pcs}) for the spin-$\frac{1}{2}$ chain of alternating Ising and Heisenberg spins.}
\label{fig5}
\end{figure}

\begin{figure}[!t]
\begin{center}
\includegraphics[width=7.5cm]{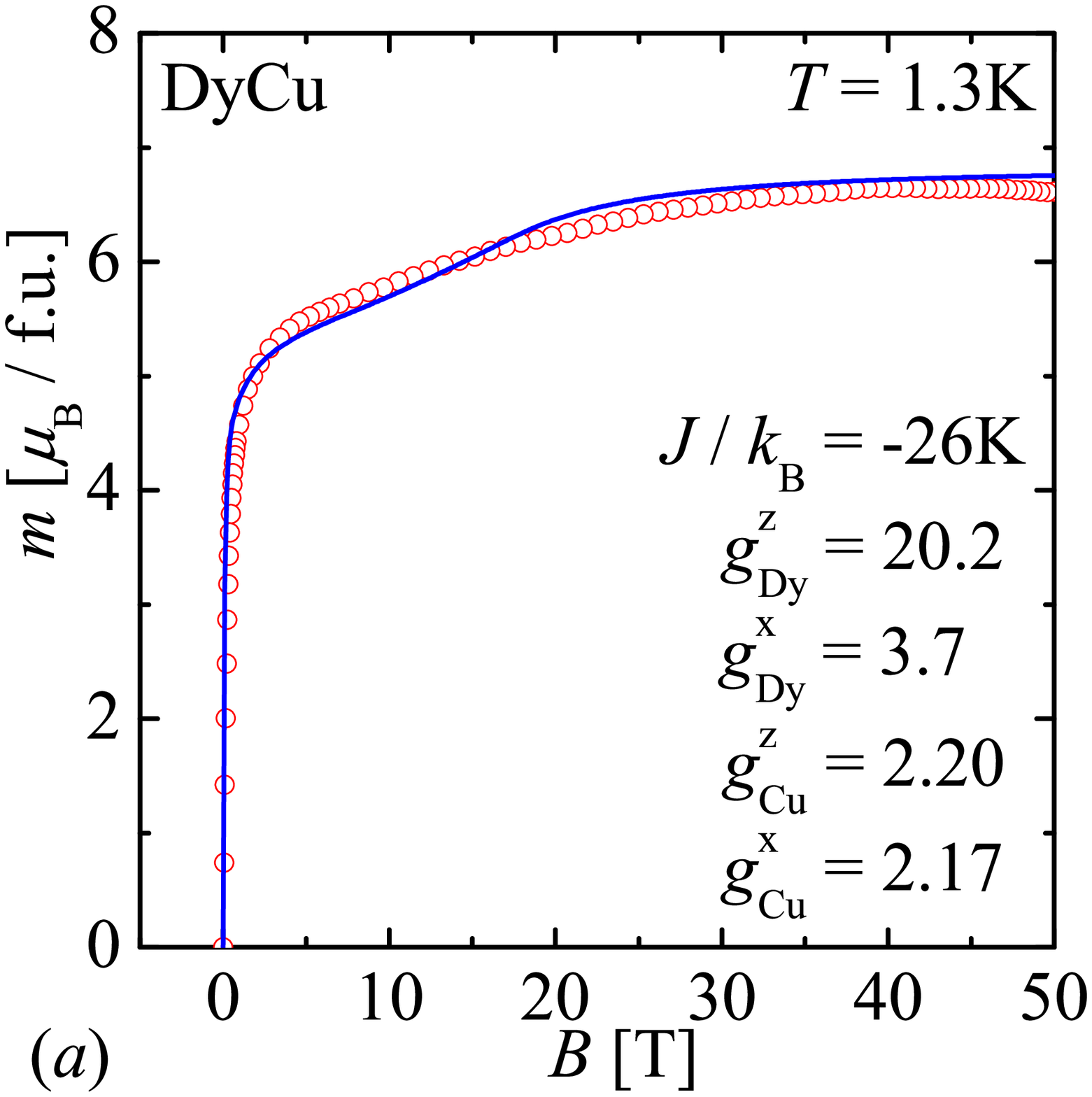}
\hspace{-2.0cm}
\includegraphics[width=7.5cm]{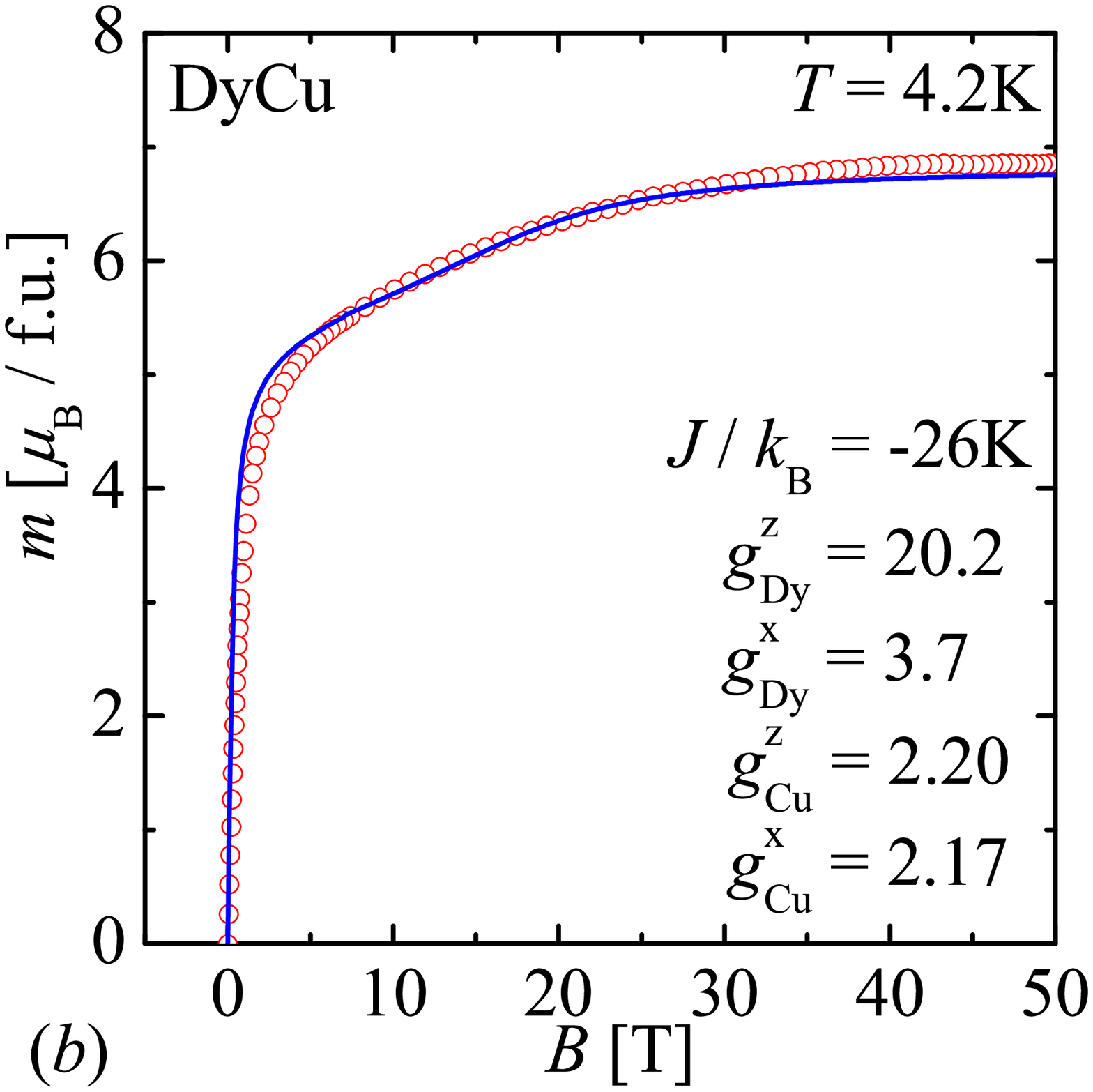}
\end{center}
\caption{(Color online) High-field magnetization data of the powder sample of DyCu at two different temperatures: (a) $T=1.3$~K; (b) $T=4.2$~K. Blue solid lines show
the best theoretical fit obtained using the formulas~(\ref{scsmfa}) and~(\ref{pcs}) for the spin-$\frac{1}{2}$ chain of alternating Ising and Heisenberg spins
with the mean-field correction for the transverse magnetization of the Ising spins.}
\label{fig6}
\end{figure}

High-field magnetization data of the powder sample of DyCu after subtracting the temperature-inde\-pendent paramagnetism are presented in figure~\ref{fig5} together with the best theoretical fit obtained by using the magnetization formulas~(\ref{scs}) and~(\ref{pcs}) derived for the ferrimagnetic spin-$\frac{1}{2}$ chain of alternating Ising and Heisenberg spins. As one can see, one generally obtains a quite plausible concordance between the recorded experimental data and the relevant theoretical predictions
except a small discrepancy observable in the low-field region, where a more abrupt change in the magnetization is theoretically predicted than the one experimentally observed.
It should be nevertheless mentioned that the determined value of Land\'e factor of Dy$^{3+}$ ion $g_1^{z} = 22.6$ is somewhat greater than theoretically expected.
This might indicate a small but non-negligible effect of the transverse component of $g$-factor, which was completely ignored for Dy$^{3+}$ ions within our exact treatment.
If the transverse component $g_1^x$ of Land\'e factor is taken into account for Dy$^{3+}$ ions at the mean-field level, the magnetization formula~(\ref{scs})
for a single-crystal sample can  be easily corrected for the contribution of the transverse magnetization of the Ising spins (Dy$^{3+}$ ions) yielding
\begin{eqnarray}
m (\theta) = (m_1^z + m_2^z) \cos \theta  + (m_1^x + m_2^x) \sin \theta
\label{scsmfa}
\end{eqnarray}
with
\begin{eqnarray}
m_1^x \equiv g_1^x \mu_{\rm B} \langle \sigma^x_i \rangle = \frac{g_1^x \mu_{\rm B} H_1^x}{\sqrt{\left(2J m_2^z + H_1^z\right)^2 + \left(H_1^x\right)^2}} \frac{1}{2} \tanh \left[\frac{\beta}{2}
\sqrt{\left(2J m_2^z + H_1^z\right)^2 + \left(H_1^x\right)^2} \right].
\label{m1x}
\end{eqnarray}
In doing so, the corrected magnetization formula~(\ref{scsmfa}) can be substituted into the formula~(\ref{pcs}) derived for the polycrystalline system in order to analyze
the magnetization curve of the powder sample of DyCu. In this way, one actually resolves the problem with a too high value of the Land\'e $g$-factor of Dy$^{3+}$ ions
as evidenced by the best fitting data set indicated in figure~\ref{fig6}. It is also quite evident from figure~\ref{fig6} that the acquired theoretical prediction
follows more accurately the relevant experimental data in the low-field region and beside this, the values of $g$-factors gained for Dy$^{3+}$ and Cu$^{2+}$ ions
from this latter fitting procedure are simultaneously much more reliable.

\section{Concluding remarks}
\label{conclusion}

The present article is devoted to an exact study of the spin-$\frac{1}{2}$ chain of alternating Ising and Heisenberg spins in the magnetic field of arbitrary spatial direction. The low-temperature magnetization curve of the investigated spin-chain model was scrupulously examined in the dependence on a spatial orientation of the applied magnetic field. It has been demonstrated that the magnetization curve becomes smoother, the intermediate plateau shrinks, and the total magnetization is reduced by quantum fluctuations as the external field deviates from the easy axis of the Ising spins. Accordingly, the powder averaging in the related polycrystalline system yields a smooth low-temperature magnetization curve, which quite closely follows the magnetization curve of a single-crystal system for a spatial orientation of the external field deviating by $\theta = 60^{\circ}$ from the easy-axis direction.

The proposed spin-chain model has been employed for interpreting high-field magnetization data of polymeric coordination compound DyCu, which provides an interesting
experimental realization of the ferrimagnetic spin-$\frac{1}{2}$ chain of regularly alternating Dy$^{3+}$ and Cu$^{2+}$ magnetic ions reasonably approximated by the notion
of Ising and Heisenberg spins, respectively. From this perspective, experimental studies performed on single-crystal and field-aligned samples of DyCu represent
a particularly challenging problem for future investigations.

\section*{Acknowledgements}
This work was partly supported by the Global COE Program (Core Research and Engineering of Advanced Materials-Interdisciplinary Education Center for Materials Science) (No.~G10) from the Ministry of Education, Culture, Sports, Science and Technology (MEXT), Japan. J.S. acknowledges warm hospitality during his stay as visiting research scholar at KYOKUGEN centre.

\ukrainianpart

\title{Феримагнітний спін-1/2 ланцюжок з почергових Гайзенбергових та Ізингових спінів у довільно орієнтованому
магнітному полі}

\author{Й. Стречка\refaddr{label1}, М. Хагівара\refaddr{label2},
Й. Ган\refaddr{label2}, T.~Кіда\refaddr{label2}, З.~Гонда\refaddr{label3}, M.~Ікеда\refaddr{label2}}
\addresses{
\addr{label1} Природничий факультет, Унiверситет iм. П.Й. Шафарика, Кошiце, Словацька республiка
\addr{label2} KYOKUGEN (Центр квантових наук і технологій),
Університет м. Осака,  560--8531 Осака, Японія
\addr{label3} Факультет функціонального матеріалознавства, Вища школа  природничих наук та інженерії, \\
Університет м. Сайтама,  338--8570 Сайтама, Японія}

\makeukrtitle

\begin{abstract}
\tolerance=3000%
Феримагнітний спін-1/2 ланцюжок, який складається з почергових Гайзенбергових та Ізингових спінів у довільно орієнтованому
магнітному полі, розв'язується точно, використовуючи перетворення повороту спінів та метод трансфер-матриці. Показано,
що низькотемпературний процес намагнічення залежить в основному від просторової орієнтації магнітного поля. Гостра сходинкоподібна
форма кривої намагніченості з помітним проміжним плато, яке з'являється у магнітному полі прикладеному вздовж напрямку легкої осі Ізингових спінів,
стає гладшою та проміжне  плато коротшає, якщо зовнішнє поле відхиляється від напрямку легкої осі. Крива намагніченості полікристалічної системи також обчислюється,
здійснюючи конфігураційне усереднення отриманої формули для намагніченості. Запропонована модель спінового ланцюжка дає розуміння намагніченості у сильних полях $3d$-$4f$
біметалічної полімеpної сполуки Dy(NO$_3$)(DMSO)$_2$Cu(opba)(DMSO)$_2$, яка допускає  цікаву експериментальну реалізацію феримагнітного ланцюжка, складеного з двох різних, але регулярно змінних спін-1/2 магнітних іонів Dy$^{3+}$ та Cu$^{2+}$, що прийнятно апроксимуються поняттями Ізингових та Гайзенбергових спінів, відповідно.

\keywords феримагнітний спіновий ланцюжок, точні результати, плато намагніченості, $3d$-$4f$ біметалічна сполука

\end{abstract}

\end{document}